\documentstyle[12pt,aasms]{article}
\lefthead{Kim et al.}
\righthead{Solar Convection}
\begin{document}
\title{Modelling of shallow and inefficient convection in the outer
layers of the Sun using realistic physics}
\author{Yong~-Cheol Kim}
\affil{Center for Solar and Space Research,
Yale Astronomy Department, Box 208101, New~Haven, CT~06520-8101}
\author{Peter A. Fox}
\affil{HAO/NCAR, P.O.Box 3000, Boulder, CO 80307-3000}
\and
\author{Sabatino Sofia, and Pierre Demarque}
\affil{Center for Solar and Space Research,
Yale Astronomy Department, Box 208101, New~Haven, CT~06520-8101}
\begin{abstract}
In an attempt to understand the properties of convective energy transport
in the solar convection zone,
a numerical model has been constructed for turbulent flows in a compressible,
radiation-coupled, non-magnetic, gravitationally stratified medium
using a realistic equation of state and realistic opacities.
The time-dependent, three-dimensional hydrodynamic equations are solved
with minimal simplifications.

The statistical information obtained from the present simulation
provides an improved understanding of solar photospheric convection.
The characteristics
of solar convection in shallow regions is parameterized and
compared with the results of Chan and Sofia's simulations of deep and
efficient convection (Chan and Sofia 1989).
 We assess the importance of
the zones of partial ionization in the simulation, and confirm that the
radiative energy transfer is negligible throughout the region except in
the uppermost scale heights of the convection zone, a region of very
high super-adiabaticity.

When the effects of partial ionization are included,
the dynamics of flows are altered significantly.   However, we
confirm the Chan and Sofia result
that kinetic energy flux is non-negligible and can have a negative value in
the convection zone.

\end{abstract}

\section{Introduction}

To investigate the mean (horizontally and time--averaged) properties of
stellar convection zones,
and to test the validity of the mixing length approximation,
Chan and Sofia initiated a numerical study of deep, efficient convection
in a compressible and stratified layer (Chan and Wolff
1982; Chan et al.\ 1982; Sofia and Chan 1984;
Chan and Sofia 1986; Chan and Sofia 1987; Chan and Sofia 1989).
In their calculations,
the effects of partial ionization and energy transfer by
radiation were ignored.
Among several important results in their papers were; 1)
the mixing length is proportional to the pressure scale height, not to
the density scale height (Chan and Sofia 1987), 2)
the kinetic energy flux is non-negligible and negative over most of the
layer (Chan and Sofia 1989; see also Cattaneo et al.\ (1991)).
 This implies that the mixing length approximation
under-estimates the enthalpy flux, the heat transfer by convective
elements, and 3)
there are simple relationships between fluctuating quantities and the mean
dynamical and thermodynamical structure which are similar to those given by the
mixing length approximation (Vitense 1953, B\"ohm-Vitense~1958;
 Chan and Sofia 1989).
Thus, certain parts of the mixing length theory were found to be valid in
the limit that the convection is deep and efficient, a condition which applies
to much of the solar convection zone.

Unfortunately, solar (and stellar)
structure models have their greatest sensitivity to the mixing length
parameter(s) in regions of partial ionization and where radiative
energy transport becomes important.  For example, Kim et al.\ (1991)
show what mixing length parameters are required when changes are
made to the equation of state, and alternate radiative opacities
are included.   Whenever opacity increases, we need to increase the
efficiency of convective energy transport by increasing the mixing length
ratio.  Note, however, that different mixing length ratios did not
cause much structural change in most of convection zone where
the $\nabla - \nabla_{ad}$ is negligible.

In the shallow layers of stars, regions of partial ionization provide an
additional driving (and perhaps damping) force for convection motions.
To date, few quantitative modeling efforts have addressed this issue
(see Rast and Toomre 1993ab, Rast et al.\ 1993).
Furthermore,
information on convection in the convective-radiative transition layers
is vital to many branches of solar physics (solar dynamo,
solar variability, solar atmosphere modeling, etc.) in addition to the
general study of stellar structure.
It is not {\it a priori} clear to what degree the models of deep convection
can represent such layers.
In this paper, we investigate whether the results from
studies of deep and efficient convection still hold in a shallow and
inefficient convective zone, where the convection becomes less efficient
and where ionization plays important role in the dynamics of
convection.

The numerical model for solar photospheric convection in a compressible,
radiation-coupled, non-magnetic, gravitationally stratified medium
is described in section 2.

By performing statistical analyses of the simulation, we determine
properties
of solar convection, and compare them with those obtained by Chan and
Sofia(1989), in section 3.
In section 4, the results of this study are summarized and discussed.

\section{Convection simulation}

\subsection{Hydrodynamical equations}

Consider a fluid moving in a known gravitational field $g$, carrying no
electric charge or electric current, and
undergoing no chemical reactions.
In a non-rotating frame of reference, the equations representing this
stratified
compressible hydrodynamical flow in three dimensions are:
\begin{eqnarray}
  {{\partial \rho} \over{\partial t}} & = & - \nabla \cdot {\bf M},
  \label{eq:cmass} \\
  {{\partial {\bf M}} \over{\partial t}} & = &
  - \nabla \cdot { \left[ {\bf M} {{\bf M} \over {\rho}}
  - {\bf{\Sigma}} + P {\bf I} \right] }
  + \rho {\bf g},
 \label{eq:cmtm} \\
  {{\partial \varepsilon } \over{\partial t}} & = &
  - \nabla \cdot \left[\left(\varepsilon  + P \right)
 { {\bf M} \over {\rho} }
  - { {\bf M} \over {\rho} } \cdot {\bf{\Sigma}} + \bf{f} \right]
  + {\bf M} \cdot {\bf g},
 \label{eq:cener}
\end{eqnarray}
\begin{eqnarray}
  {\varepsilon} & = & e+{\frac{1}{2}}\rho {\bf v}^2,
  \qquad{\rm Total \ Energy} \label{eq:totenrg} \\
  P & = & P\left(e,\rho,\mu\right),\label{eq:pressure}
 \qquad{\rm Pressure}\\
  T & = & T\left(e,\rho,\mu\right),\label{eq:temperature}
  \qquad{\rm Temperature}\\
  {\bf{\Sigma}}  & = & 2\bar{\mu} {\bf{\sigma}} +
        \lambda\left(\nabla\cdot {\bf v} \right){\bf I},
\qquad{\rm Viscous \ Stress
        \ Tensor} \label{eq:viscstress}
\end{eqnarray}
where $\rho$ is the density, ${\bf M}$ is the mass flux vector ($\rho {\bf
v}$),
$e$ is the specific internal energy,
 $\mu$ is the mean molecular weight,
${\bf g}$ is the gravitational acceleration,
${\bf I}$ is the identity tensor, and
${\bf{\sigma}}$ is the usual rate of strain tensor ($ {\sigma}_{ij} =
{{\left({\nabla}_{i} {v}_{j} + {\nabla}_{j} {v}_{i} \right)} \over {2}}$).

The goal of the direct simulations of stellar convection is to solve the
equations governing compressible fluid dynamics in a star with the fewest
possible approximations.  Some approximations, however, are unavoidable.
One such assumption concerns the resolution of the simulation which
models flows with large Reynolds numbers and large Rayleigh numbers.

It is not
possible with currently available computer power to encompass all scales
which are present in stellar convection.   In most stellar
convection zones, the effective Rayleigh numbers (Ra) and the Reynolds
numbers (Re) are very large.    Since the ratio between a pressure scale
height and the Kolmogorov microscale, below which the flow really becomes
smooth, is of order ${Re}^{3 \over 4} \approx {10}^{7.5}$, ${Re}^{9 \over
4} \approx {10}^{22.5}$ grid points are required to resolve the motions in
all scales.   Even if advances in computer capabilities made such a
simulation feasible, we may not be interested in such
fine resolution, since the smallest scales do not participate in the heat
transport.

In any case, current hydrodynamic simulations
calculate only the larger scales of the convective flow, and use
some prescription to take into account the smaller, subgrid scales.
The simplest way is to choose a fixed, larger value of the viscosity to
represent the effect of the Reynolds stresses due to the subgrid scales,
assuming that the scale is within an inertial cascade range
in the local turbulence spectrum.   Since our grid spacings
and the actual Reynolds number were not uniform, however,
we employed a variable,
also larger, viscosity which raises the local effective grid Reynolds
number to order unity everywhere (Ramshaw~1979).
Instead of using the gas viscosity values, the forms of $\bar{\mu}$ and
$\lambda$ in equation (\ref{eq:viscstress})
are chosen to represent the effects of the Reynolds stresses on unresolved
scales (Smagorinsky 1963):
\begin{eqnarray}
  \bar{\mu} & = & \rho \left( c_{\mu} \Delta \right)^2
                 \left( 2{\bf{\sigma}}:{\bf{\sigma}}\right)^{\frac{1}{2}},
                 \qquad{\rm Dynamic \ Viscosity}
 \label{eq:viscosity}
\end{eqnarray}
where the Deardorff coefficient $c_\mu=0.2$ was used.
Ignoring bulk viscosity, the quantity $\lambda$ was
taken to be ${-2/3\bar{\mu}}$ (Stoke's hypothesis).

The total heat flux ${\bf F}$ is made up of
$  {\bf F} = {\bf F}_{e} + {\bf F}_{k} + {\bf F}_{v} + {\bf f} $
where the enthalpy, kinetic energy, viscous and diffusive fluxes
respectively are:
\begin{eqnarray}
  {\bf F}_{e} & = & \left(e + P\right){ {\bf M} \over {\rho} }
  \label{eq:fluxep}\\
  {\bf F}_{k} & = & \left(\frac{1}{2} \rho V^2\right){\bf V}
  \label{eq:fluxk}\\
  {\bf F}_{v} & = & - {\bf V} \cdot {\bf{\Sigma}}
  \label{eq:fluxv}\\
  {\bf f} & =  & - K_T \nabla T + K_p \nabla P ,
  \label{eq:fluxsgs}
\end{eqnarray}
and thermal conductivities are:
\begin{eqnarray}
  K_T & = &  \frac{\mu}{Pr} C_p + \frac{4acT^3}{3k\rho}
  \label{eq:therm1a} \\
  K_p & = &   \frac{\mu}{Pr} C_p \nabla_{ad} \frac{T}{P},
  \label{eq:therm2}
\end{eqnarray}
where $c_p$ is the specific heat at constant pressure,
$\nabla_a$ is the adiabatic
gradient $\equiv {{\partial \ln T} \over {\partial \ln P}} |_{ad}$,
and $Pr = \nu/\kappa$ is
the Prandtl number ($\nu$ is the kinematic viscosity  and
$\kappa$ is the thermal diffusivity).
Radiative transfer is treated with the diffusion approximation, where $k$
is the opacity, $a$ is the Boltzmann constant and $c$ is the speed of light.
When the mean free path of a photon is much shorter than the depth of the
zone, the radiative diffusion treatment is a good approximation
for the equation of radiative transfer.   Therefore, in the most of our
calculation domain, the approximation is expected to be valid.
 At the top, however, the photon mean free path is $ {1
\over 10}$ of the depth.   It is not clear that the approximation
is still valid in the hoped-for degree.
--- The photon mean free path (${1 \over {\kappa \rho}}$)
is about $9 \times 10^6 cm$.
The top of the calculation domain locates at the depth of $ 9 \times
10^7 cm$.   The top of the solar model is defined where the density is
$10^{-10} g cm^{-3}$.---

\subsection{The numerical solution; ADISM}

 The equations governing the fluid motion form a system of
coupled non-linear partial
differential equations.   Although this system is based on the very
simple physical principles of conservation of mass (\ref{eq:cmass}),
momentum (\ref{eq:cmtm}), and energy (\ref{eq:cener}),
analytical solutions to this system of equations can only be found under some
very specific and often unrealistic assumptions (special symmetry, small
perturbations, etc.).     Therefore, for physically realistic solutions,
this system of equations must be solved numerically.

To study large scale convection in the Sun, the system of
the hydrodynamic equations was solved using the `Alternating Direction
Implicit on Staggered Mesh' method (ADISM; Chan and Wolff~1982).
This method was accurate to second order even for non-uniform zoning
(staggered mesh), free from the time step restrictions for stability
 (implicit method),
and efficient for solving multi-dimensional hydrodynamics equations
(Alternating Direction method).   The code used for this research was
a three-dimensional and non-magnetic
version of the code Fox developed (Fox et al.\ 1991),
using Chan's GENTRX code (Chan and Wolff~1982).

While the grid spacing was uniform in the horizontal direction, the spacing
in the vertical direction was chosen to be non-uniform, so that higher
resolution can be obtained near the top of the calculation domain where
scale heights are small.  The $i^{th}$ layer is located at
\begin{eqnarray}
r_i & = & {{r_2} \over {1+  {{(r_2 - r_1)} \over {Z_G r_1}}
\lbrace \exp \lbrack{{N_r - i} \over {N_r - 1}}  \ln(1+Z_G)\rbrack -1
 \rbrace}},
  \label{eq:zgrd}
\end{eqnarray}
where $r_1, r_2$ are the bottom and top vertical coordinates, respectively,
$N_r$ is the number of grids in the vertical direction, and $Z_G$ is the
controlling parameter of the vertical grid distribution (Chan and Sofia
1986).

To control numerical truncation errors caused by the nonuniform vertical
gridding, a coordinate transformation has been introduced in such a way
that the mapped grids are uniform, and
all the quantities are made dimensionless so that the initial values
of the density, temperature, and pressure at the top are equal to 1.

\subsection{Initial conditions}

A solar model is constructed using the Yale stellar evolution code
(Guenther et al.\ 1989, Kim et al.\ 1991, Guenther et al.\ 1992) with
solar abundances given by Anders-Grevesse (1989) and opacities from the
Los Alamos Opacity Library (LAOL) tables (Huebner et al.\ 1977).

The equation of state is divided into two separate regions and an intermediate
transition region.  In the outer region, the equation of state routine
determines particle densities by solving the Saha equation for the single
ionization state of $H$ and heavier elements, and the single and double
ionization states of $He$.  Perturbations due to Coulomb interactions and
excited states in bound systems are neglected.  In the inner regions, the
elements are assumed to be fully ionized.   The partially degenerate, partially
relativistic case is handled with an iteration method. In the transition
region, the interior and exterior formulation are weighted with a ramp function
and averaged.

The solar $He$ mass fraction, $Y$ (which cannot be determined by observation),
together with the mixing length parameter, $\alpha$, are free parameters in
stellar model construction.   As is usual with the given model physics,
$\alpha$ and $Y$ were adjusted in such a way that a solar model has the solar
radius and the solar luminosity at the solar age (Guenther 1989). The
characteristics of the model are summarized in table~(\ref{tbl:model_star}).
\begin{table}
%\centering
%\begin{tabular}{lc} \hline \hline
%Parameters                                   & Sun     \\ \hline
%Mass $({{M} \over {{M}_{\odot}}})$            & 1.00  \\
%Luminosity $(\log {{L} \over {{L}_{\odot}}})$ & 0.00  \\
%Radius $(\log {{R} \over {{R}_{\odot}}})$     & 0.00  \\
%Effective temperature$(\log {T}_{eff})$      & 3.76   \\
%Age (Gyr)                                    & 4.50   \\
%Mixing length ratio  & 1.230 \\
%Weight fraction of hydrogen (X) & 0.7071 \\
%Weight fraction of all heavy elements (Z) & 0.0188 \\
%Mixture of heavy elements  & Anders-Grevesse(1989) \\ \hline \hline
%\end{tabular}
\caption{Solar model \label{tbl:model_star}}
\end{table}

The stratifications of the internal energy, the density, and the gravity of the
photospheric convection zone in the model were then taken as the initial
condition for the hydrodynamic simulation. This ensures consistency between the
shallow regions where the photospheric convection was simulated and the deeper
regions of the stellar structure. The initial condition used for the simulation
is shown in figure \ref{fig:suninit}.
\begin{figure}[e]
%\centerline{\hbox{
%{\psfig{figure=../guesswhat/sunprho.ps,height=10cm}}
%{\psfig{figure=../guesswhat/suntrho.ps,height=10cm}}
%}}
\caption[The initial stratification of the solar convection simulation]
{The hydrostatic structure constructed using Yale stellar evolution code was
used as the initial condition for the hydrodynamic simulation of the
Sun.
	 \label{fig:suninit}}
\end{figure}
Since the hydrodynamic equations are solved in nondimensional form, the
scaling factors are obtained from the surface level of the solar
models and are listed in table (\ref{tbl:model_init}).
\begin{table}
\caption{Scaling factors associated with the initial condition.
	 \label{tbl:model_init}}
%\end{minipage}
\end{table}

A small kinematic perturbation was applied to initiate motion:
\begin{eqnarray}
  M_x & = & V_0 \rho \sin\theta_x
\left( -\cos\theta_z \sin\theta_y + \sin \theta_z \cos\theta_y \right),
  \label{eq:initvx} \\
  M_z & = & -V_0 \rho \cos\theta_x \sin\theta_z
\left( -\cos\theta_y + \sin\theta_y \right),
  \label{eq:initvz} \\
  M_y & = & V_0 \rho \sin\theta_y
\left( \cos\theta_x \sin\theta_z - \sin\theta_x \cos\theta_z \right),
  \label{eq:initvy}
\end{eqnarray}
where
$$
\theta_i ={{N_i (grid~position) \pi} \over {number~of~grid~spacings}}.
$$
$i$ represents the spatial coordinates,
$V_0$ is the amplitude of the initial perturbation, and
$N_i$ controls the number of cells in the initial perturbation.

\subsection{Boundary conditions}
The computational domain was a rectangular box with
impenetrable, stress-free boundaries at the top and bottom.  A constant heat
flux, which was calculated from the hydrostatic model structure,
was imposed at the bottom.   The input flux is shown as $F\over F_T$
in table (\ref{tbl:model_init}).   Thus:
\begin{equation}
{\bf V}\cdot{\bf n} = 0, \qquad
\frac{\partial}{\partial n}\left({\bf V}\times{\bf n}\right) = 0,
\label{eq:sidebc}
\end{equation}
\begin{equation}
F_z = F_{bottom} = {\hbox{\rm constant (at bottom only)}},
\end{equation}
\begin{equation}
e  = e_{top} = {\hbox{\rm constant (at top only)}}.
\end{equation}
Periodic boundary conditions for all variables were used in the horizontal
direction.

\subsection{Simulation specifications}
The hydrodynamic equations (conservation of mass (\ref{eq:cmass}),
momentum (\ref{eq:cmtm}), and energy (\ref{eq:cener})), together with
the same equation of state and opacities which were used in the initial model
constructions, were solved in time, until a converged statistically
steady-state solution was reached.
To control the time step of the simulations, a dimensionless number was
defined: $$ N_{CFL} = {{\Delta t C_s} \over {\Delta r}}, $$ where $C_s$
is the sound speed at the top, ${\Delta r}$ is the minimum grid spacing,
and $\Delta t$ is the time step.  Since our mesh system is nonuniform in
the sense that we can obtain higher resolution at the top, $\Delta r$ in
this case is usually $\Delta z$ at the top.

Further conditions for the simulations are summarized in Tables
(\ref{tbl:sim}) and (\ref{tbl:initpur}).
\begin{table}
%\centering
%\begin{minipage}{8.0cm}
%\begin{tabular}{lc} \hline \hline
%Simulation &  Sun \\ \hline
%Side boundary  & periodic \\
%Number of grids ($x \times y \times z$) &
%$32 \times 32 \times 32$ \\
%vertical size ($z/R_T$)   & $0.996 \sim 1.0$ \\
%Horizontal size ($x/R_T$) & $0.992 \sim 1.0$ \\
%Horizontal size ($y/R_T$) & $0.992 \sim 1.0$ \\
%  $Z_G$  & 3.0  \\
%$N_{CFL}$ & 0.2 \\
%Vertical size in PSH \footnote{Pressure Scale Height}
% & 6.0558 \\
%\hline \hline
%\end{tabular}
\caption{The simulation specification.
	 \label{tbl:sim}}
%\end{minipage}
\end{table}
%$\beta$ in equation (\ref{eq:adi}) & 0.6 \\
%
\begin{table}
%\centering
%\begin{tabular}{lcl} \hline \hline
%Parameter & Sun  & Description \\ \hline
%$V_0$ &  0.5  & the amplitude of the initial perturbation \\
%$N_x$ & 2.0& the number of eddys in x direction \\
%$N_z$ & 1.0&the number of eddys in z direction \\
%$N_y$ & 2.0&the number of eddys in y direction \\
%\hline \hline
%\end{tabular}
\caption{The initial perturbation.
	 \label{tbl:initpur}}
\end{table}

To determine whether a calculation has `relaxed', the time
history of the horizontally averaged surface heat flux and the maximum RMS
velocity in the domain were monitored.

In figure \ref{fig:spden}, \ref{fig:sptem}, and \ref{fig:spvel}, density and
temperature contours and velocity fields of the simulation are shown at a time
after the statistically steady-state was reached. To visualize the
three-dimensional flow pattern, contours on one x-z plane,  one y-z plane, and
three different levels of x-y plane are shown.

The left-top panel in figure \ref{fig:spvel} shows the flow pattern which
resembles the solar granules and intergranule regions.  The topology change
with depth is shown at the left three panels in figure \ref{fig:spvel}. The
relative density fluctuations and the relative temperature fluctuations are
higher at the top.

The tight correlation between the radial velocity and the temperature and the
density fluctuations at the top found in other studies (Chan and Sofia 1986,
1989) is also evident in these simulations. For example, the anti-correlation
between the radial velocity and the density fluctuation is -0.8 or greater.

While these figures are useful in obtaining the gross character of the
simulated convection, more physical information can only be extracted from
the simulations using statistical quantities.
For example, in the case of stars, the observed spectral absorption line
asymmetries only provides temporally and spatially
integrated information on the surface convective motions. Even for solar
convection, which is the only case where time-resolved information can be
obtained, the overall characteristics, not detailed temporal variations,
are of interest in this paper.
%The gross properties of the convection can be
%obtained from the spatial and temporal average of the convective behavior.
{}From the detailed analysis of the statistical information, the validity of
prescriptions of convective energy transport, such as the mixing length
approximation, can be tested.

%\include{cplts2}
% Sun periodic
\begin{figure}[e]
%\centerline{\hbox{
%{\vbox{
%{\psfig{figure=../guesswhat/sp27den.ps,height=7cm}}
%{\psfig{figure=../guesswhat/sp16den.ps,height=7cm}}
%{\psfig{figure=../guesswhat/sp5den.ps,height=7cm}}
%}}
%{\vbox{
%{\psfig{figure=../guesswhat/spyden.ps,height=7cm}}
%{\psfig{figure=../guesswhat/spxden.ps,height=7cm}}
%}}
%}}
\caption[The density contour]
{The density contours.  This snapshot of
the solar convection simulation is at time step 94000.
At the left column from top to bottom, density contours
are on x-y plane at z=27, 16, and 5, respectively.
The contour on x-z plane at y=16 grid is shown at right-top panel.
The right-bottom panel shows the contour on y-z plane at x=16.
\label{fig:spden}}
\end{figure}
\begin{figure}[e]
%\centerline{\hbox{
%{\vbox{
%{\psfig{figure=../guesswhat/sp27tem.ps,height=7cm}}
%{\psfig{figure=../guesswhat/sp16tem.ps,height=7cm}}
%{\psfig{figure=../guesswhat/sp5tem.ps,height=7cm}}
%}}
%{\vbox{
%{\psfig{figure=../guesswhat/spytem.ps,height=7cm}}
%{\psfig{figure=../guesswhat/spxtem.ps,height=7cm}}
%}}
%}}
\caption[The temperature contours]
{The temperature contours at the same time step as the figure
\ref{fig:spden}.
At the left column from top to bottom, the temperature contours
are on x-y plane at z=27, 16, and 5, respectively.
The contour on x-z plane at y=16 grid is shown at right-top panel.
The right-bottom panel shows the contour on y-z plane at x=16.
\label{fig:sptem}}
\end{figure}
\begin{figure}[e]
%\centerline{\hbox{
%{\vbox{
%{\psfig{figure=../guesswhat/sp27vel.ps,height=7cm}}
%{\psfig{figure=../guesswhat/sp16vel.ps,height=7cm}}
%{\psfig{figure=../guesswhat/sp5vel.ps,height=7cm}}
%}}
%{\vbox{
%{\psfig{figure=../guesswhat/spyvel.ps,height=7cm}}
%{\psfig{figure=../guesswhat/spxvel.ps,height=7cm}}
%}}
%}}
\caption[The velocity fields]
{The velocity fields at the same time step as the figure
\ref{fig:spden}.
At the left column from top to bottom, the velocity fields
are on x-y plane at z=27, 16, and 5, respectively.
The projection of the velocity vectors on the x-y plane
and the contours of $V_z$ are shown.  The solid-line and the broken-line
contours represent positive and negative radial velocity, respectively.

The fields on x-z plane at y=16 grid is shown at right-top panel.
The projection of the velocity vectors on the x-z plane
and the contours of $V_y$ are shown.  The solid-line and the broken-line
contours represent positive and negative $V_y$, respectively.

The right-bottom panel shows the fields on y-z plane at x=16.
The projection of the velocity vectors on the y-z plane
and the contours of $V_x$ are shown.  The solid-line and the broken-line
contours represent positive and negative $V_x$, respectively.
\label{fig:spvel}}
\end{figure}
%
%\end{document}

%\include{result2}
\section{Statistical properties}

Statistical averages of the simulation are calculated to study the general
properties of solar convection.
The simulation carried out corresponds to
$2.08 \times 10^5$ seconds of solar convection.
The horizontal and temporal averages of the
quantities of interest were calculated during a period of about 16
turnover times ($7.11 \times 10^4$ seconds in the simulation).
The characteristics of convection were parameterized by relationships
suggested both by mixing length theory and the results of Chan and
Sofia (1989), and are
summarized in table \ref{tbl:stat_rel}.
These results were calculated using  the middle 4 pressure scale heights  of
the layer, where the mean vertical mass flux
$\langle \rho V_z \rangle  \leq 1 \times 10^{-4}$, implying that the
mean distribution of the
fluid was not undergoing substantial re-adjustment.

For a quantity $a$, ${\langle}a{\rangle}$ denotes the combined
horizontal and temporal mean, $a'$ denotes the deviation from the mean, and
$a''$ denotes the root mean square fluctuation from the mean.  The correlation
coefficient of two quantities $a_1$ and $a_2$ is expressed
as $C[a_1, a_2]$:
$$
 C[a_1, a_2] = {{{\langle}a_1 a_2{\rangle}} \over
   {\sqrt{{\langle}{a_1}^2{\rangle}}\sqrt{{\langle}{a_2}^2{\rangle}}}}.
$$
All quantities in the tables are scaled values. (See
table \ref{tbl:model_init}).
\begin{table}
\caption{Approximate relationships obtained from the numerical
simulations.
	 \label{tbl:stat_rel}}
\end{table}

In Table \ref{tbl:stat_rel}, our results are compared with those of
Chan and Sofia (1989).
At this stage, it is necessary to restate some difference between the
physical conditions in each study. Since their interest was
in deep, efficient convection
(Chan and Wolff 1982; Chan et al.\ 1982; Sofia and Chan 1984;
Chan and Sofia 1986; Chan and Sofia 1987; Chan and Sofia 1989),
the effects of ionization and energy transfer by
radiation were ignored. Since our interest is in the shallower
layers, these effects are present in our simulation and thus are expected to
produce different parameterizations.
For the present simulation,
it turned out that the effect of radiative energy transport was quite
small throughout the calculation domain, except in the uppermost scale height.
Therefore, in most regions in the solar convection zone, ignoring
the coupling between radiation and convection seems to be a reasonable
approximation.  (More discussion regarding the effect of radiative energy
transfer can be found later together with the distribution of fluxes.)

Chan and Sofia (1989) used a unit constant mean molecular weight and,
by changing the ratio of specific heats, they
calculated convection of gases with different ionization stages.
Since our simulations used an equation of state in which
the ionization of $H$ and $He$ was calculated based on the local
thermodynamic conditions, the mean molecular weight varied
temporally and spatially.  The temporal and horizontal average of the mean
molecular weight is shown in figure \ref{fig:lnpm}.

Finally, because our initial structure was obtained from a solar model, the
simulation domain contained a highly super-adiabatic region as shown in
figure \ref{fig:lnpdel}, near 12.5 in logarithmic pressure. The Chan
and Sofia simulations used an initial polytropic structure model and
thus this feature was absent.
\begin{figure}[e]
%\centerline{\psfig{figure=mmt.ps,width=14.0cm}}
\caption[The temporal and horizontal average of the mean molecular weight]
{The temporal and horizontal average of the mean molecular weight
calculated from the simulation.
\label{fig:lnpm}}
\end{figure}
\begin{figure}[e]
%\centerline{\psfig{figure=deldel.ps,width=14.0cm}}
\caption[The temporal and horizontal average of the super adiabatic gradient]
{The temporal and horizontal average of ($\nabla -\nabla_{ad}$)
calculated from the simulation is represented by the crosses.   The solid
line represents the initial stratification.
The stratification of the simulation is steeper than that of the initial model.
\label{fig:lnpdel}}
\end{figure}

{}From Table (\ref{tbl:stat_rel}), a few characteristic differences can be
found.   The horizontal velocity fluctuation is comparable to the
vertical velocity fluctuation, except near the top and bottom where the
boundary condition enforces $V_z''$ to be zero and the horizontal velocities
are large. This seemingly isotropic velocity fluctuation, shown
in figure \ref{fig:relvfl}, did not exist in the
calculations of Chan and Sofia (1989).
\begin{figure}[e]
%\centerline{\psfig{figure=velrms.ps,width=14.0cm}}
\caption[The velocity fluctuation]
{Root mean square velocity fluctuations versus depth.
In this figure, like all others, the depth is specified by
$\ln(P/P_{top})$, where $P_{top}$ is the pressure at the top of the
thermally relaxed fluid.
The solid, the dashed, and the dotted lines show $V_z''$, $V_x''$, and $V_y''$
respectively.
\label{fig:relvfl}}
\end{figure}

The relative fluctuations of the thermodynamic variables
${\rho''/{\langle}{\rho}{\rangle}}$, ${T''/{\langle}T{\rangle}}$, and
${P''/{\langle}P{\rangle}}$, shown in figure \ref{fig:relflt}, seem to
indicate that the ratio between the three fluctuation quantities changes
in the highly super-adiabatic region.   It is, however, difficult to judge
whether the different characteristics are from the degree of super-adiabaticity
or the top boundary condition.   As shown in figure \ref{fig:lnpdel}, the
relaxed solution has the pressure enhanced at the top.
Even in the regions where
$(\nabla - \nabla_{ad})$ was close to zero, the ratios between the
fluctuating quantities are different from that of Chan and Sofia (1989).
In addition,
${P''/{\langle}P{\rangle}}$ and ${\rho''/{\langle}{\rho}{\rangle}}$ are
larger than ${T''/{\langle}T{\rangle}}$.  Therefore, the contribution of
the pressure term in the entropy fluctuation
$\delta S = C_p ( \delta \ln T - \nabla_{ad}\ \delta \ln P )$
is not negligible, in contrast to their result.
\begin{figure}[e]
%\centerline{\psfig{figure=relflc.ps,width=14.0cm}}
\caption[The relative fluctuation of the thermodynamic variables]
{Distributions of the relative fluctuations of the thermodynamic variables.
The solid, the dashed, and the dotted lines are the
temperature, pressure, and density fluctuations, respectively.
\label{fig:relflt}}
\end{figure}

In the present simulations, it was found that
the relative pressure fluctuation was proportional to the square of
the Mach number.   The coefficient of the relation between
${P''/{\langle}P{\rangle}}$ and ${\mu V''^2/{\langle}T{\rangle}}$, however,
was different (Table \ref{tbl:stat_rel}).

The relation between ${P''}$ and ${\rho {V_z''}^2}$ was derived from the
relation between ${P''/{\langle}P{\rangle}}$ and
${\mu V''^2/{\langle}T{\rangle}}$
and the relative ratio between the fluctuations of
the thermodynamic parameters. Since the relative ratios among the fluctuating
quantities are different in the highly super-adiabatic region, as shown in
figure \ref{fig:relflt}, the correlation is not as tight as in the calculation
of Chan and Sofia (1989).

The mean values of the correlation coefficients between pairs of parameters
are shown in Table \ref{tbl:stat_rel}
 and in figure \ref{fig:cpos} and \ref{fig:cneg}.
\begin{figure}[e]
%\centerline{\psfig{figure=corrp.ps,width=14.0cm}}
\caption[Correlation between fluctuating quantities I]
{Correlation between fluctuating quantities.
The solid, the dashed, the dash-dot, and the dotted lines are
$C[T', S']$, $C[P', T']$, $C[V_z, T']$, and $C[V_z, S']$, respectively.
\label{fig:cpos}}
\end{figure}
\begin{figure}[e]
%\centerline{\psfig{figure=corrn.ps,width=14.0cm}}
\caption[Correlation between fluctuating quantities II]
{Correlation between fluctuating quantities.
The solid, the dashed, and the dotted lines are
$C[\rho', S']$, $C[V_z, \rho']$, and $C[\rho', T']$, respectively.
\label{fig:cneg}}
\end{figure}
While others correlations in Table \ref{tbl:stat_rel}
are close to what was found for deep efficient convection,
$C[\rho', T']$, $C[P', T']$, and  $C[P', T']$ are noticeably different.

Further relations for the covariance of $V_z$ with the thermodynamical
variables are calculated to compare with those of Chan and Sofia (1989).
The covariance
${{{\langle}V_z \rho'{\rangle}} /
                {{\langle}V_z{\rangle}{\langle}\rho{\rangle}}} = -1 $
implies simply mass conservation, ${{\langle}\rho V_z {\rangle}}=0$.
Therefore,  this indicates that the mean distribution of the fluid in the
simulation is no longer undergoing substantial adjustment.
The covariance,
${{{\langle}V_z P {\rangle}} /
                {{\langle}V_z{\rangle}{\langle}P{\rangle}}} \approx 1.75 $,
which implies
${{{\langle}V_z P' {\rangle}} /
                {{\langle}V_z{\rangle}{\langle}P{\rangle}}} \approx 0.75 $,
and the covariance
${{{\langle}V_z T' {\rangle}} /
                {{\langle}V_z{\rangle}{\langle}T{\rangle}}}\approx 0.85$,
implies that
$P'/P$, and $T'/T$ are not negligible in the convective flow.

The characteristic differences in the properties of convection in this
research and in Chan and Sofia (1989) can be summarized as follows:  unlike
their deep efficient convection, the velocity fluctuations were close to
isotropic and the pressure and density fluctuations are larger than the
temperature fluctuations;   the correlations among $T'$, $P'$, and $\rho'$
found in this research are quite different from those in Chan and Sofia
(1989).  It appears that the inclusion of ionization was responsible for
these differences.   In a perfect gas, the temperature change causes the
density change.   When there is a strong correlation between these two
variables, the correlation of these parameters with pressure must be low.
This is what we observe from Chan and Sofia's simulation.   On the other
hand, when the temperature change is not directly connected to the density
change, because of ionization, the correlation  of these parameters
with pressure must be stronger.  Since ionization changes the number of
light particles (electrons), the pressure change will be larger than the
density change. This simple explanation helps one to understand the
differences in the characteristics of convection in this research and in
Chan and Sofia (1989).

The differences between convection in the two studies can also be shown in
another way. From their deep efficient convection simulations, Chan and Sofia
(1987, 1989) showed that the vertical correlation length of vertical velocity
and temperature deviation are scaled by the pressure scale height, not by the
density scale height. In their research, they calculated several cases with
different ratios of specific heats, $\gamma$.  The practical meaning of
different $\gamma$'s is different mean molecular weights.   Therefore, each
calculation with a different $\gamma$ can `simulate' the convection of gases at
different ionization stages. When the vertical two-point correlations of $V_z$
from several simulations with different $\gamma$'s were plotted together, they
scaled with pressure scale height, not with density scale height. When the
partial ionization process is coupled with the convection calculation, the
situation is changed.  In figure \ref{fig:corvv} and \ref{fig:cortt} the
vertical, two-point correlations at five different depths are plotted together.
 As shown in figure \ref{fig:lnpm}, since the average mean molecular weight
varies as a function of depth, it practically means each symbol corresponds to
a case with different $\gamma$.
\begin{figure}[e]
%\centerline{\hbox{
%{\psfig{figure=pvv.ps,height=10cm}}
%{\psfig{figure=rhovv.ps,height=10cm}}
%}}
\caption[Two-point correlation of vertical velocity]
{Distributions of the two-point correlation of vertical velocities are plotted
against the logarithmic pressure (HWHM $\sim 1.5$) and the logarithmic density
(HWHM $\sim 1.2$) for different depths. \label{fig:corvv}}
\end{figure}
\begin{figure}[e]
%\centerline{\hbox{
%{\psfig{figure=ptt.ps,height=10cm}}
%{\psfig{figure=rhott.ps,height=10cm}}
%}}
\caption[Two-point correlation of temperature fluctuation]
{Distributions of two-point correlation of temperature fluctuation are plotted
against the logarithmic pressure and the logarithmic density for different
depths.
\label{fig:cortt}}
\end{figure}
The vertical two-point correlation of $V_z$ can be scaled with the density
scale height as well as with the pressure scale height (cf. Chan and Sofia
1987).  The vertical correlation of $T'$, however, can not be scaled with
pressure scale height, nor density scale height.

Yet another difference is apparent in the distribution of energy fluxes.
\begin{figure}[e]
%\centerline{{\psfig{figure=sumagn.ps,width=14.0cm}}}
\caption[Distribution of heat fluxes]{Distribution of heat fluxes.
The total flux
(the thick solid line) is the summation of the enthalpy flux (the thin solid
line), the kinetic energy flux (the dashed line), the viscous flux (the
dotted line), the radiative flux (the dot-dash line),
and the diffusive flux (the dot-dot-dot-dash line).
\label{fig:ecompnt}}
\end{figure}
The distribution, shown in figure \ref{fig:ecompnt} basically confirms the
study of Chan and Sofia (1989).  The kinetic energy flux was not negligible in
the convection zone, and it was negative. This important result holds in
shallow region convection as well as in deep efficient convection. The enthalpy
flux is larger than the total flux. Recall that in the mixing length
approximation, the total flux is carried by two components only, radiation and
convection, the latter is equivalent to the enthalpy flux and thus is
under-estimated (see the discussion in Lydon et al.\ 1992). This result casts
further doubts on the mixing length approximation in the shallow convective
regions of a star.

Compared with Chan and Sofia (1989), however, the diffusive flux is much
larger.    When zones of partial ionization overlap with the convection zone, a
certain amount of the energy flux is taken away for ionization so that the same
amount of energy need not be transferred by means of convection.  Practically
speaking, $C_p$ changes in the diffusion term in the energy equation, equation
(\ref{eq:cener}).  An increase in $C_p$ means an increase in the thermal
conductivity (see equation (\ref{eq:fluxsgs}), (\ref{eq:therm1a}), and
(\ref{eq:therm2})). This larger diffusive action may explain the seemingly
isotropic velocity fluctuation discussed earlier. Chan and Sofia (1989) found
the vertical velocity fluctuations are larger than the horizontal velocity
fluctuations in their convective flows.

The effect of the radiative energy transport in the convection zone is apparent
in figure \ref{fig:ecompnt}.  The radiative energy flux (the dot-dash line) is
small throughout the region except at the top one pressure scale height.   This
result confirms the assumption used for the simulations of deep efficient
convection by Chan and Sofia (1989): The radiative flux can be assumed to be
negligible for deep efficient convection.

One must remember that this simulation has a rigid top and bottom boundary
condition. Therefore, flows which impinge upon the boundary have to diverge. As
a result, the $V_z$ is smaller close to the top and bottom boundary. This
causes the enthalpy flux and  the kinetic energy flux to be small at the
bottom.   Only the diffusive flux can be increased to carry the flux out at the
bottom. At the top, the radiative flux takes over the transfer of the energy
flux.

There are two additional comments for figure \ref{fig:ecompnt}. Firstly, at the
top the radiative flux exceeds the total flux, and the sub-grid scale diffusive
flux becomes negative (i.e. an entropy inversion).   This indicate that the
`mean' structure in the region has become stable to convection.  This
conclusion must be viewed cautiously since this effect happens at the very top
of the vertical grid and may be an un-physical artifact of the top boundary
condition.

Secondly, at the bottom the kinetic energy flux is positive, whence through out
most of the region it is negative.  Because of the rigid bottom boundary, the
characteristic flow pattern is modified so that down-ward flows can not but
diverge and larger horizontal velocities are induced. As a result, the kinetic
energy flux can not be negative.

In summary, our research confirms one assumption which Chan and Sofia employed
for their study: that radiation can be de-coupled from convection in regions
where the motions are deep and efficient. The contribution of the radiative
flux in our simulation became significant only at the top one pressure scale
height. The partial ionization process, however, should not be de-coupled from
convection in the shallow layers. As shown in figure \ref{fig:relvfl} through
\ref{fig:ecompnt} and summarized in table \ref{tbl:stat_rel}, the effects of
partial ionization is not negligible.

\section{Summary}
        Three-dimensional hydrodynamic simulations for the shallow layers of
the solar convection zone have been performed in order to probe the general
properties of convective energy transport. Our simulations were made using a
realistic equation of state and opacities and allowed for radiative energy
transport in the diffusion approximation. To ensure the overall consistency of
the structure, the initial model for the convection simulation was taken from a
detailed solar structure model constructed using the Yale stellar evolution
code.

 From the detailed analysis of the simulation, the characteristics of solar
convection in shallow regions have been parameterized and compared with that of
Chan and Sofia's deep, efficient convection simulations (Chan and Sofia 1989).
{}From the  differences between the statistical properties obtained, two
important points were drawn.  First, the transition from the radiative mode to
the convective mode of energy transport occurs at a very thin layer at the top
of the convection zone.     The contribution of energy transport by radiation
in the simulation was negligible throughout most of the domain except within
the top one pressure scale height.   Therefore, it is safe to de-couple the
radiation from convection simulation except in super-adiabatic regions.

Secondly, the thermodynamic effects of partial ionization on the dynamics of
the flow were significant and led to different statistical properties compared
to the study of Chan and Sofia which used a fully ionized gas with different
specific heats. Therefore if the emphasis of a simulation is on detailed
comparison with solar surface features (for example), every effort should be
made to include a realistic equation of state.

Comparison with three of the results of Chan and Sofia deserve to be pointed
out especially. Firstly, unlike the result of Chan and Sofia (1987, 1989), our
study shows that the pressure scale height is not a preferable scaling factor
over the density scale height in the shallow convective layers. Secondly, in
agreement with Chan and Sofia(1989), we find that the kinetic energy flux in
the upper convective zone is not negligible, and that it is mostly downward
directed. As a result, the temperature gradient which the mixing length
approximation prescribes, will not be appropriate for that part of the
convection zone. Finally, they have shown that the mixing length approximation
is valid in the limit of deep and efficient convection; we will test this
result in a forth-coming paper.

\acknowledgments
We wish to acknowledge the invaluable contributions of K. L. Chan
in the current research.
This research was supported in part by AFOSR91-0053 and NASA grant
NAGW-2531 to Yale University.   Computer time was provided by the
National Center for Atmospheric Research.

\end{document}